\documentclass[conference]{IEEEtran}
\IEEEoverridecommandlockouts
\usepackage{cite}
\usepackage{amsmath,amssymb,amsfonts}
\usepackage{svg}

\usepackage{graphicx}
\usepackage{textcomp}
\usepackage{url}
\usepackage{xcolor}
 \usepackage{booktabs}
 \usepackage{algorithm}
\usepackage{algpseudocode}
 \usepackage{array} 
\usepackage{tabularx}

\def\BibTeX{{\rm B\kern-.05em{\sc i\kern-.025em b}\kern-.08em
    T\kern-.1667em\lower.7ex\hbox{E}\kern-.125emX}}

\begin{document}

\twocolumn[
\begin{@twocolumnfalse}

\begin{center}
{\color{blue}
\large This paper is accepted for presentation at the IWCMC 2026 Conference, June 1--6, 2026, Shanghai, China.
}
\vspace{1em}
\end{center}

\title{Risk-Aware and Stable Edge Server Selection Under Network Latency SLOs}

\author{
\IEEEauthorblockA{
Mohan Liyanage, Arnova Abdullah, Eldiyar Zhantileuov, Rolf Schuster \\
University of Applied Sciences and Arts, Computer Science Department, Dortmund, Germany. \\
mohan.liyanage@fh-dortmund.de
}
}

\maketitle

\end{@twocolumnfalse}
]

\begin{abstract}
 We present a lightweight and interpretable decision framework for dynamic edge server selection in latency-critical applications that explicitly accounts for tail risk and switching stability. Each candidate server is characterised by predictive mean and uncertainty summaries of network latency, which are used to estimate the risk of service-level objective (SLO) violations and to guide selection. Risk is evaluated using a tight Normal approximation complemented by a conservative Cantelli bound, while percentile-based scoring coupled with hysteresis stabilizes decisions and suppresses oscillatory switching under short-lived network fluctuations.

Experimental results on a multi-server edge testbed with a strict SLO of
$\tau = 0.5$\,s show that the proposed approach reduces the deadline-miss rate
from 39\% to 34\% compared to a mean-only baseline, while reducing switching frequency from 46\% to 5.5\% (88\% reduction) and maintaining sub-SLO average
latency (0.429 s).
These results demonstrate that explicit risk evaluation combined with stability-preserving control enables practical and robust adaptive server selection in dynamic edge environments.

\end{abstract}

\section{Introduction}
\label{sec:intro}

Edge computing is a key paradigm for latency-sensitive applications that require
real-time processing close to data sources, such as intelligent transportation
systems, cooperative perception, and roadside video analytics. In these
applications, strict end-to-end (E2E) latency constraints must be satisfied to
maintain Quality of Service (QoS), safety, and user experience, making dynamic
edge server selection a critical challenge.

Recent work has proposed lightweight models for predicting network latency in
edge environments. However, selecting the server with the lowest predicted mean
latency alone is often insufficient in practice. Mean-only selection ignores
latency variability and predictive uncertainty in shared and time-varying
networks, and thus may underestimate the risk of service-level objective (SLO)
violations. Moreover, naive reactive policies are highly sensitive to short-lived
fluctuations, causing oscillatory switching between servers and incurring
handover overheads such as connection resets, transient packet loss, and
warm-up delays. These effects can increase tail latency and undermine
reliability even when average delay remains acceptable.

Our earlier work showed that E2E latency in edge networks exhibits nonlinear
growth near saturation and proposed a rational delay model to capture this
behavior~\cite{paper1}. We later introduced reliability scoring to improve
robustness in server selection~\cite{paper2}. However, mean-driven selection can
still remain overly reactive and may fail to capture variance-driven tail risk.
This motivates the need for a decision layer that converts uncertain latency
predictions into reliable and stable server-selection actions under strict SLOs.

In this paper, we formulate dynamic edge server selection as a
\emph{risk-aware and stability-preserving decision problem}. Each candidate
server is characterized by predictive summaries of network latency, namely a
mean and an uncertainty estimate derived from recent observations. We propose a
lightweight two-stage framework. First, SLO-violation risk is evaluated using a
Normal surrogate for interpretability together with a Cantelli bound as a
conservative safeguard against distributional mismatch. Second, server selection
is stabilized using percentile-based scoring with hysteresis, so that switching
occurs only when a candidate provides sustained and meaningful improvement over
the current server.

The proposed approach is designed for practical edge environments where only
end-to-end observable performance metrics are available and internal server
states may be inaccessible. Thus, unlike queue-aware or learning-based methods,
it does not rely on detailed infrastructure visibility, offline training, or
complex online optimization. This makes the framework lightweight,
interpretable, and suitable for black-box deployment settings.

The novelty of this work lies not in introducing a new latency predictor, but
in designing a lightweight decision layer that transforms uncertain delay
summaries into explicit SLO-risk estimates and stable server-selection actions.
The main contributions are: 1) a hybrid Normal--Cantelli risk evaluation
mechanism for SLO-aware server ranking; 2) a percentile-based hysteresis policy
to suppress oscillatory switching; and 3) an experimental validation on a
multi-server edge testbed showing lower deadline-miss rate and substantially
reduced switching compared with a mean-only baseline.
 \section{Related Work}
\label{sec:relwo}

\subsection{Risk Evaluation in Networking and Real-Time Systems}
Evaluating the risk of delay violations is a long-standing problem in network
performance analysis and real-time systems. Normal approximations are commonly
used to derive interpretable percentile-based guarantees~\cite{harchol2013performance},
but empirical network delays often exhibit burstiness and heavy-tailed behavior,
which can limit the reliability of purely Gaussian models. To address such
uncertainty, distribution-free concentration inequalities, including Chebyshev’s
and Cantelli’s bounds, are widely used to provide conservative tail
guarantees~\cite{grimmett2020probability,mitzenmacher2017probability,buttazzo1997hard,stewart2005performance}.
These tools are especially useful in safety-critical settings, where optimistic statistical estimates are often complemented by conservative safeguards.

\subsection{Risk-Aware MEC and Edge Server Selection}
In Mobile Edge Computing (MEC) systems, dynamic server selection under uncertain
network conditions has received increasing attention. Traditional heuristics,
such as nearest-server or least-load selection, expose the trade-off between
latency and resource utilization under dynamic workloads~\cite{10710290}. More
recent work has explored reliability-aware server selection using predictive
summaries and heuristic rules to improve robustness under fluctuating network
conditions~\cite{paper2}. Learning-based approaches have also been proposed for
adaptive edge selection, task offloading, and SLO-aware resource management in
dynamic MEC environments~\cite{kang2024slo,zhang2025tail}.

Probabilistic bounds, including Cantelli-type inequalities, have further been
used in wireless and edge networking for risk-aware decision making, such as
resource allocation and access control~\cite{11194191}. However, existing
approaches typically emphasize either risk estimation or adaptive control, while
paying limited attention to switching stability. As a result, server selection
may remain vulnerable to oscillatory behavior under short-lived fluctuations.

In contrast, our work focuses on a lightweight \emph{decision layer} that
explicitly transforms mean and uncertainty summaries into SLO-risk estimates and
combines them with percentile-based hysteresis. This enables risk-aware and
stability-preserving edge server selection using only end-to-end observable
network metrics, without requiring internal queue states, infrastructure-level
coordination, or training-based control.

\subsection{Qualitative Comparison with Existing Approaches}
Table~\ref{tab:qualitative_comparison} contrasts our framework with representative
classes of server selection methods. Our contribution complements existing
approaches by providing an interpretable, low-overhead decision layer that
explicitly evaluates tail risk and suppresses oscillatory switching under partial
observability.


\begin{table}[t]
\centering
\caption{Qualitative comparison of server selection approaches}
\label{tab:qualitative_comparison}
\footnotesize
\setlength{\tabcolsep}{4pt}
\begin{tabularx}{\columnwidth}{l c c c c}
\toprule
\textbf{Method} & \textbf{Risk} & \textbf{Stability} & \textbf{Observability} & \textbf{Training} \\
\midrule
Mean-only              & No       & No       & End-to-end only     & No  \\
Queue-aware / Lyapunov & Partial  & No       & Internal state      & No  \\
RL-based offloading    & Implicit & Implicit & Often extensive     & Yes \\
\textbf{This work}     & \textbf{Yes} & \textbf{Yes} & \textbf{End-to-end only} & \textbf{No} \\
\bottomrule
\end{tabularx}
\end{table}

Unlike learning-based or queue-aware approaches, this work focuses on
risk-aware decision control using only end-to-end observable latency statistics,
explicitly addressing switching stability, which is largely overlooked in prior
server selection frameworks.

\section{Methodology and System Model}
\label{sec:methodology}

This section formalizes the network-centric server-selection problem and the
risk-evaluation framework used to compare candidate edge servers. The goal is to
estimate the risk of violating a strict latency service-level objective (SLO)
under uncertainty, prior to applying the stability control introduced in
Section~\ref{sec:Decision_Policy}.

\subsection{System Model and Predictive Summaries}
We consider an edge-assisted application in which a client continuously offloads
data frames to one of $K$ candidate edge servers
$\mathcal{S}=\{S_1,S_2,\ldots,S_K\}$. The dominant performance bottleneck is the
network-induced delay along the client-to-server path, including transmission,
propagation, and queueing effects.

Let $d_i(t)$ denote the observed network latency associated with server $S_i$ at
decision step $t$. For each server, the decision layer maintains predictive
summaries over a sliding window of the most recent $W$ observations:
\[
\mu_i(t)=\frac{1}{W}\sum_{u=t-W+1}^{t} d_i(u),
\]
\[
\sigma_i(t)=
\sqrt{\frac{1}{W-1}\sum_{u=t-W+1}^{t}\left(d_i(u)-\mu_i(t)\right)^2}.
\]
During the initial warm-up period, when fewer than $W$ samples are available,
the summaries are computed from the available history only. Delay samples are
aggregated per frame before updating the window, which reduces sensitivity to
transient jitter and measurement noise.

Using these summaries, the network latency of server $S_i$ is approximated by a
Gaussian surrogate
\[
N_i(t)\approx \mathcal{N}\!\left(\mu_i(t),\sigma_i^2(t)\right),
\]
where $\mu_i(t)$ is the predicted mean latency and $\sigma_i(t)$ captures
uncertainty due to traffic variability, contention, and partial observability.
The QoS requirement is expressed as a strict latency SLO:
\[
N_i(t)\le \tau,\qquad \tau=500~\mathrm{ms}.
\]

\subsection{Risk Evaluation and Percentile Scoring}
Rather than relying only on mean latency, we explicitly quantify the risk that
server $S_i$ violates the SLO. Given the predictive summary
$(\mu_i,\sigma_i)$, we use two complementary risk measures.

First, under the Gaussian surrogate, the SLO-violation probability is estimated
as
\[
p_i^{\mathrm{Norm}}(\tau)=
1-\Phi\!\left(\frac{\tau-\mu_i}{\sigma_i}\right),
\]
where $\Phi(\cdot)$ is the standard Normal cumulative distribution function.

Second, to guard against distributional mismatch and heavier-tailed behavior, we compute the one-sided Cantelli bound
\[
p_i^{\mathrm{Cant}}(\tau)=
\frac{1}{1+\left(\frac{\tau-\mu_i}{\sigma_i}\right)^2}.
\]
The Normal estimate provides an interpretable and often tighter approximation,
while the Cantelli term acts as a conservative safeguard when the Gaussian assumption is less accurate.

\begin{figure}[t]
    \centering
    \includegraphics[width=0.9\linewidth]{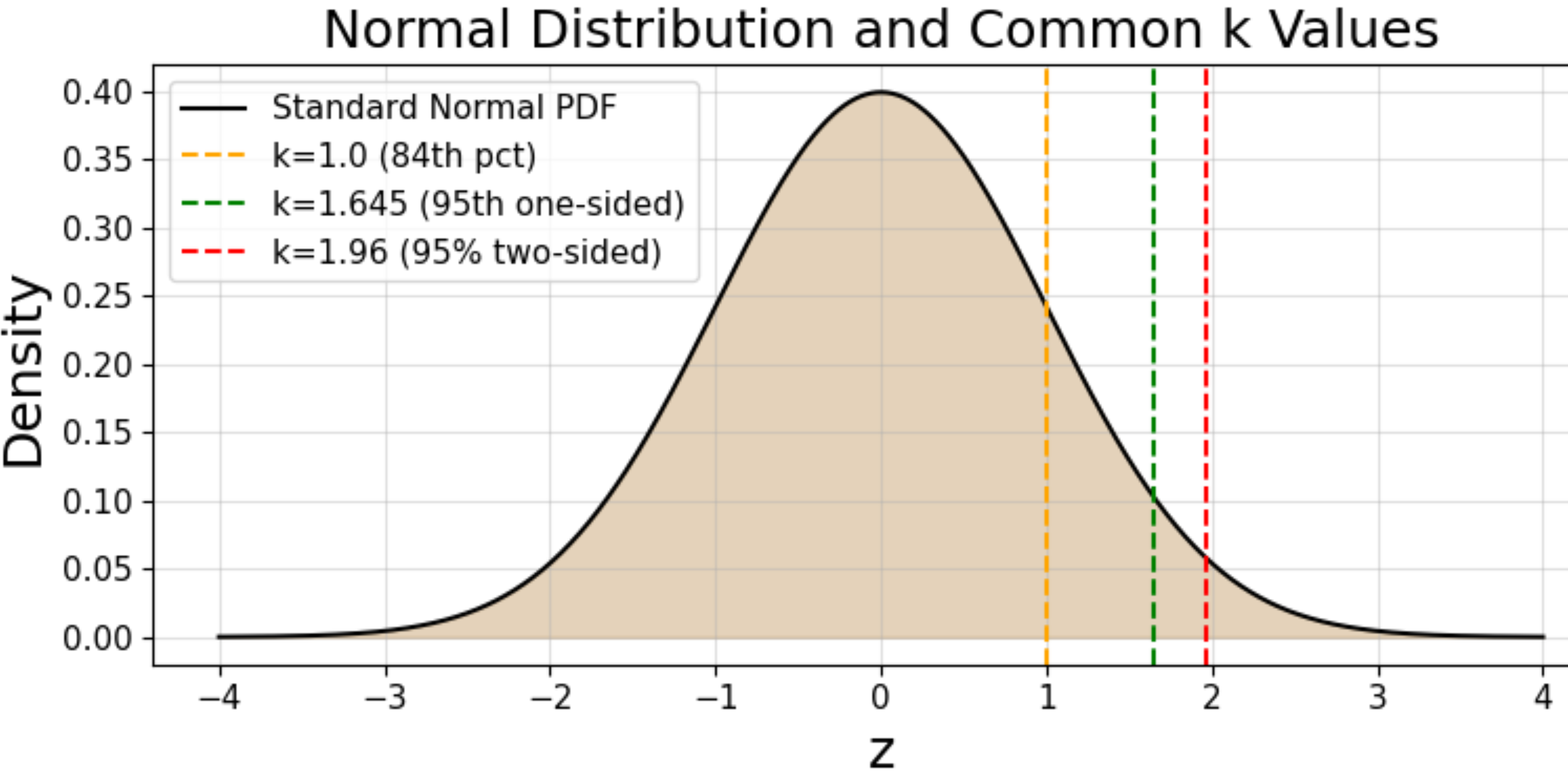}
    \caption{Interpretation of the risk-aversion parameter $k$ under a Normal
    latency surrogate. The mean delay $\mu$ corresponds to the 50th percentile,
    while $\mu+k\sigma$ maps to higher percentiles (e.g., $k=1$ corresponds to the 84th
    percentile and $k=1.645$ corresponds to the one-sided 95th percentile). Increasing $k$
    yields more conservative server selection.}
    \label{fig:k_interpretation}
\end{figure}

In addition to feasibility testing, we use a percentile-based score to rank
candidate servers:
\[
\mathrm{Score}_i(k)=\mu_i+k\sigma_i,
\]
where $k\ge 0$ is a risk-aversion parameter. Larger values of $k$ place more
weight on uncertainty and therefore favor more conservative server choices. Under
the Gaussian surrogate, $\mu_i+k\sigma_i$ corresponds to an upper-percentile
latency estimate, as illustrated in Fig.~\ref{fig:k_interpretation}. Lower
scores indicate more attractive candidates and serve as input to the stability
control mechanism described in Section~\ref{sec:Decision_Policy}.

\subsection{Hybrid Risk Evaluation Algorithm}
Algorithm~\ref{alg:hybrid_risk} identifies the lowest-risk feasible server. It
first evaluates both risk measures for every candidate server, then ranks
servers by the Normal-based violation probability, and finally accepts the best
candidate only if both the Normal estimate and the Cantelli safeguard remain
below a predefined tolerance $\epsilon$.

\begin{algorithm}[t]
\caption{Hybrid Risk Evaluation}
\label{alg:hybrid_risk}
\begin{algorithmic}[1]
\Require Predicted latency summaries $\{(\mu_i,\sigma_i)\}$, SLO threshold $\tau$, risk tolerance $\epsilon$
\Ensure Lowest-risk feasible server $S_{\mathrm{best}}$
\ForAll{servers $S_i \in \mathcal{S}$}
    \State Compute $p_i^{\mathrm{Norm}}(\tau)$ and $p_i^{\mathrm{Cant}}(\tau)$
\EndFor
\State $j \leftarrow \arg\min_i p_i^{\mathrm{Norm}}(\tau)$
\If{$p_j^{\mathrm{Norm}}(\tau)<\epsilon$ \textbf{and} $p_j^{\mathrm{Cant}}(\tau)<\epsilon$}
    \State $S_{\mathrm{best}} \leftarrow S_j$
\Else
    \State $S_{\mathrm{best}} \leftarrow \emptyset$
\EndIf
\State \Return $S_{\mathrm{best}}$
\end{algorithmic}
\end{algorithm}

\subsection{Assumptions and Limitations}
The proposed framework relies on three assumptions. First, latency is assumed to be locally stationary over short observation windows, which enables real-time updating of $(\mu_i,\sigma_i)$ from recent measurements. Second, the Gaussian surrogate is used as a lightweight approximation for percentile-based reasoning, while the Cantelli bound provides a conservative fallback under bursty or heavy-tailed behavior. Third, the framework is designed for black-box edge settings in which only end-to-end latency observations are available to the decision layer.

These assumptions enable a lightweight and interpretable decision mechanism, but they may smooth out more complex latency behavior under extreme congestion or non-stationary operating conditions. Nevertheless, the combined
Normal--Cantelli formulation provides a practical balance between analytical
simplicity, robustness, and real-time applicability.

\subsection{Partial Observability and Black-Box Servers}

The proposed framework does not require access to internal scheduling states,
queue occupancies, or resource utilization of edge servers. Instead, it relies exclusively on end-to-end observable quantities, namely measured network delays and predictive summaries derived from these measurements.

Such black-box operation is common in production edge systems, where servers may be operated by third parties or belong to federated infrastructures with limited introspection capabilities~\cite{beyer2016site}. In these settings, decision-making must be based on externally observable performance indicators rather than internal state information.

When network behavior is only partially observable, predictive uncertainty
naturally increases. In the proposed framework, this increase is explicitly
reflected in the estimated variance $\sigma_i$ and directly influences decision making through percentile-based scoring and conservative Cantelli bounds. As a result, higher uncertainty leads to more conservative server selection rather than overly optimistic or brittle decisions. This behavior is intentional and aligns with the requirements of risk-sensitive edge applications operating under incomplete information.
\section{Decision Policy and Stability Control}
\label{sec:Decision_Policy}

The objective of the decision policy is to dynamically select an edge server
that minimizes the risk of violating the latency SLO while avoiding excessive
switching under time-varying network conditions. To this end, we separate
\emph{risk feasibility} from \emph{switch execution}: Algorithm~\ref{alg:hybrid_risk}
identifies feasible low-risk candidates, while a hysteresis mechanism determines
whether switching is justified.

\subsection{Why Switching Stability Matters in Edge Systems}
\label{subsec:switching_stability}

In dynamic edge environments, short-term fluctuations in network conditions can
cause risk estimates to vary rapidly, even when candidate servers have similar
long-term performance. Reacting to such fluctuations too aggressively may lead
to oscillatory switching, which introduces handover overheads such as connection
warm-up delays, transient packet loss, and disruption of application-level
state. These effects can inflate tail latency and undermine SLO compliance.
Therefore, risk-aware ranking alone is insufficient in practice, and a
stability-preserving mechanism is needed to ensure that switching occurs only
when a candidate provides a sustained and meaningful improvement.

\subsection{Risk-Aware Server Selection}
At each decision step, the client evaluates all candidate servers using the
predictive summaries and risk measures defined in
Section~\ref{sec:methodology}. Algorithm~\ref{alg:hybrid_risk} returns the
lowest-risk feasible server, denoted by $S_{\mathrm{best}}$, if both the
Normal-based estimate and the Cantelli bound remain below the tolerance
$\epsilon$. If no feasible candidate is found, the client remains on the
currently selected server.

This feasibility test prevents switching to servers whose predicted average
latency may appear attractive but whose uncertainty implies elevated
SLO-violation risk.

\subsection{Stability via Percentile-Based Hysteresis}
Risk-aware ranking alone can still lead to oscillatory switching when competing
servers exhibit similar performance and short-lived fluctuations. To suppress
such behavior, switching decisions are filtered through a percentile-based
hysteresis rule.

Let $S_{\mathrm{curr}}$ denote the currently selected server and let
$\mathrm{Score}_i(k)=\mu_i+k\sigma_i$ be the percentile-based score defined in
Section~\ref{sec:methodology}. Among all feasible candidates, the server with
the lowest score is selected as the challenger:
\[
j=\arg\min_i \mathrm{Score}_i(k).
\]
A switch from $S_{\mathrm{curr}}$ to $S_j$ is executed only if the challenger
offers a relative improvement of at least $\Delta$ for $N$ consecutive decision
steps, i.e.,
\[
\mathrm{Score}_{\mathrm{curr}}(k)-\mathrm{Score}_j(k)
\ge \Delta \cdot \mathrm{Score}_{\mathrm{curr}}(k).
\]
Here, $\Delta>0$ is the minimum improvement threshold and $N$ is the dwell
window. Together, these parameters determine the trade-off between
responsiveness and stability: smaller values react faster to changes, whereas
larger values suppress oscillations more aggressively.

\begin{algorithm}[t]
\caption{Percentile-Based Hysteresis Control}
\label{alg:hysteresis}
\begin{algorithmic}[1]
\Require Feasible candidate set $\mathcal{F}$, percentile scores $\{\mathrm{Score}_i(k)\}$, current server $S_{\mathrm{curr}}$, improvement threshold $\Delta$, dwell window $N$, persistent counter $c$
\Ensure Updated server selection and counter $(S_{\mathrm{curr}}, c)$
\If{$\mathcal{F}=\emptyset$}
    \State $c \leftarrow 0$
    \State \Return $(S_{\mathrm{curr}}, c)$
\EndIf
\State $j \leftarrow \arg\min_{S_i \in \mathcal{F}} \mathrm{Score}_i(k)$
\If{$S_j \neq S_{\mathrm{curr}}$}
    \If{$\mathrm{Score}_{\mathrm{curr}}(k)-\mathrm{Score}_j(k)\ge \Delta \cdot \mathrm{Score}_{\mathrm{curr}}(k)$}
        \State $c \leftarrow c+1$
        \If{$c \ge N$}
            \State $S_{\mathrm{curr}} \leftarrow S_j$
            \State $c \leftarrow 0$
        \EndIf
    \Else
        \State $c \leftarrow 0$
    \EndIf
\Else
    \State $c \leftarrow 0$
\EndIf
\State \Return $(S_{\mathrm{curr}}, c)$
\end{algorithmic}
\end{algorithm}

\subsection{Discussion}
The overall policy balances three objectives:
\begin{itemize}
    \item \textbf{Reliability}, by rejecting servers with high predicted
    SLO-violation risk;
    \item \textbf{Efficiency}, by favoring low percentile-based delay scores;
    \item \textbf{Stability}, by requiring sustained improvement before
    switching.
\end{itemize}

This separation between feasibility and switching control is important in
practice. A server may be statistically preferable at a single decision step,
yet not sufficiently better to justify migration overheads such as connection
warm-up, transient packet loss, or state disruption. The hysteresis layer
therefore acts as a guard against unnecessary handovers.

In the experiments, the parameters $\Delta$ and $N$ are kept fixed to provide a
clear and reproducible operating point. While adaptive tuning is possible, it is
outside the scope of this work.
\section{Experimental Evaluation}
\label{sec:exp-res}

We evaluate the proposed risk-aware and stability-preserving server-selection
framework using a real edge testbed and a broader-scale replay-based scenario.
The evaluation focuses on three aspects: reliability, measured by deadline-miss
rate (DMR) under a strict latency SLO; efficiency, measured by latency; and
stability, measured by server switching frequency.

\subsection{Experimental Setup, Policies, and Metrics}
The real testbed consists of three candidate servers ($S_1$, $S_2$, $S_3$) and
a single client continuously offloading frames from a traffic sign detection
application. Each frame corresponds to one network request with a strict
end-to-end latency SLO of $\tau=0.5$\,s. For each server, the decision layer
maintains predictive summaries $(\mu_i,\sigma_i)$ over sliding windows of recent
latency observations, as described in Section~\ref{sec:methodology}.

All policies are executed on the Client Application Management Station (CAMS).
We compare three policies: 1) a \emph{mean-only baseline} that selects the
server with the lowest predicted mean delay $\mu_i$; 2) \emph{Hybrid Risk
Evaluation} (Algorithm~\ref{alg:hybrid_risk}), which selects the lowest-risk
feasible server based on the Normal estimate and Cantelli safeguard; and
3) \emph{Percentile-Based Hysteresis} (Algorithm~\ref{alg:hysteresis}), which
applies stability control over the feasible candidates using
$\mathrm{Score}_i(k)=\mu_i+k\sigma_i$.

Reliability is measured by the DMR, defined as the fraction of frames whose
latency exceeds $\tau$. Efficiency is assessed using mean latency and, where
relevant, the 95th-percentile delay (P95). Stability is quantified by the
fraction of decision steps that result in a server switch. Unless stated
otherwise, we use $\tau=0.5$\,s, $\epsilon=0.15$, $k=1.645$, $\Delta=0.05$, and
a dwell window of $N=5$ frames.

\subsection{Main Results}
\begin{figure}[htbp]
  \centering
  \includegraphics[width=1\linewidth]{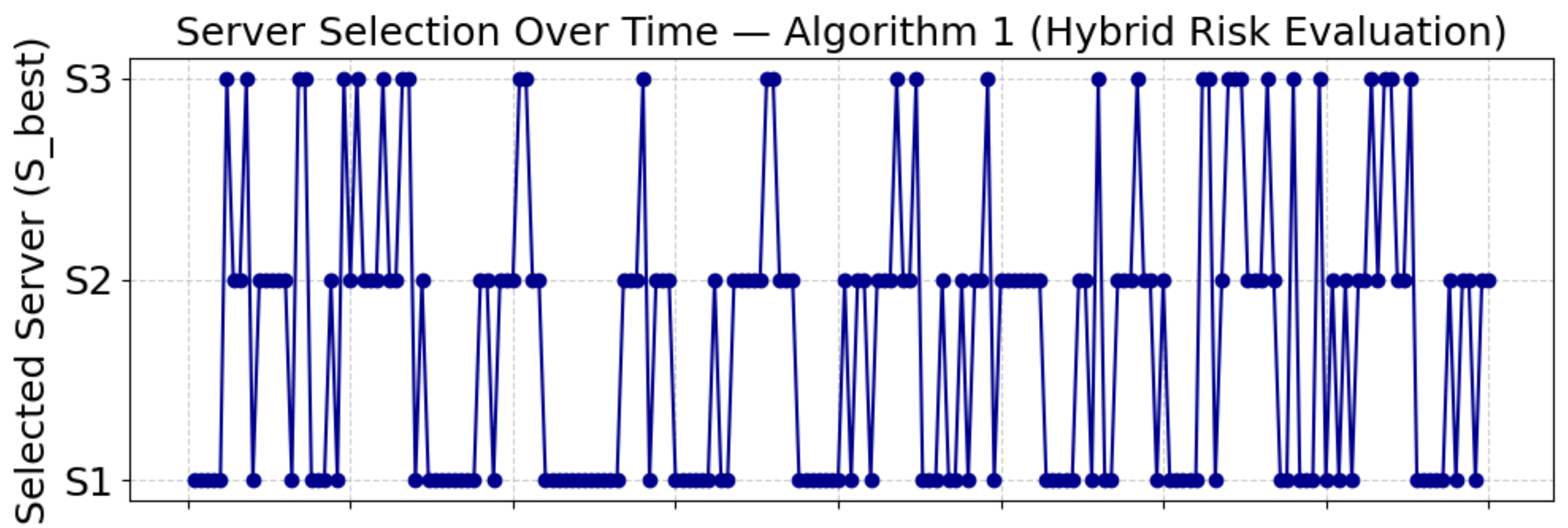}
  \caption{Server selection over time under Algorithm~\ref{alg:hybrid_risk}. The
  selected server changes frequently, indicating oscillatory switching and high
  sensitivity to short-lived fluctuations in estimated risk.}
  \label{fig:alg1_switching}
\end{figure}

\begin{figure}[htbp]
  \centering
  \includegraphics[width=0.9\linewidth]{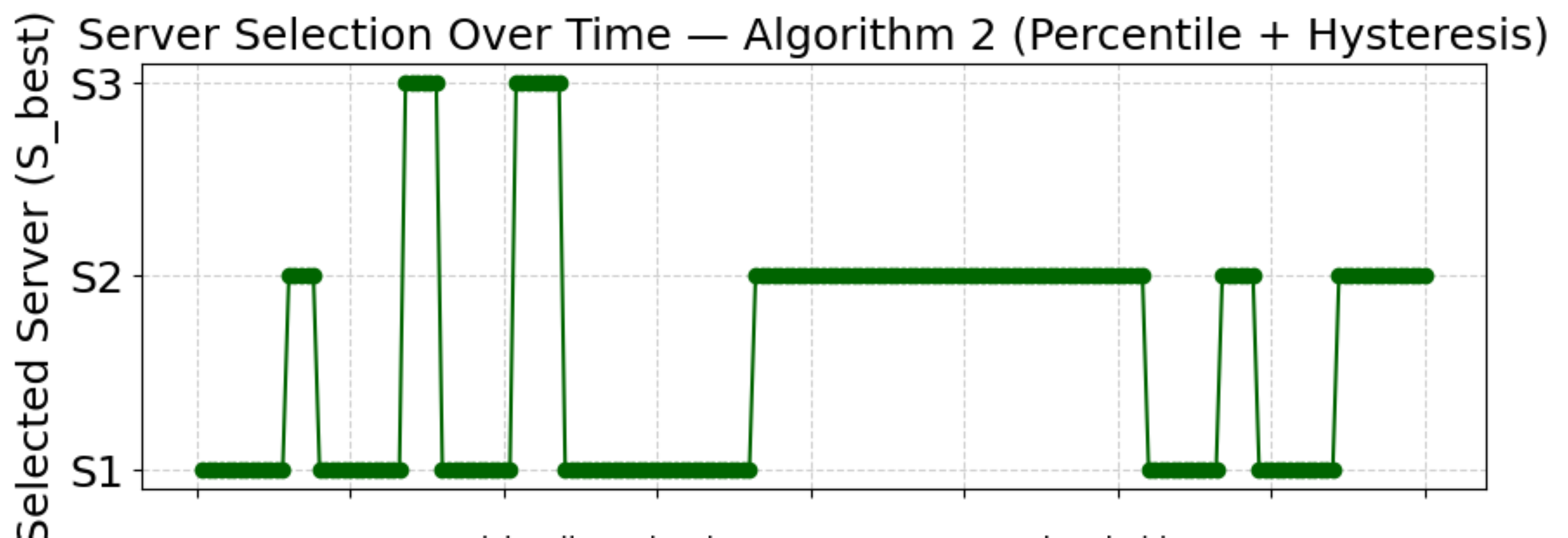}
  \caption{Server selection over time under Algorithm~\ref{alg:hysteresis}. The
  hysteresis layer suppresses oscillations by requiring sustained improvement
  before switching, resulting in more stable server assignments.}
  \label{fig:alg2_switching}
\end{figure}

Table~\ref{tab:evaluation_results} and
Figs.~\ref{fig:alg1_switching}--\ref{fig:alg2_switching} summarize the results on
the real testbed. The mean-only baseline achieves the highest DMR, while
Algorithm~2 achieves the lowest average delay, the same reduced DMR as
Algorithm~1, and substantially lower switching frequency. In particular, both
risk-aware policies reduce DMR from 39\% to 34\%, showing that accounting for
uncertainty improves reliability under the SLO.

However, Algorithm~\ref{alg:hybrid_risk} alone remains highly reactive and
switches frequently. Adding hysteresis preserves the same DMR while reducing the
average delay to 0.429~s and the switching rate from 89.5\% to 5.5\%, indicating
that stability control is essential under fluctuating network conditions.

\begin{table}[htpb]
\centering
\caption{Comparison of average delay, deadline-miss rate, and switching frequency}
\label{tab:evaluation_results}
\renewcommand{\arraystretch}{1.1}
\setlength{\tabcolsep}{3pt}
\begin{tabular}{
    l
    >{\centering\arraybackslash}p{1.3cm}
    >{\centering\arraybackslash}p{1.2cm}
    >{\centering\arraybackslash}p{1.5cm}
}
\toprule
\textbf{Algorithm} & \textbf{Avg Delay (s)} & \textbf{DMR (\%)} & \textbf{Switches (\%)} \\
\midrule
Mean-only Baseline               & 0.448 & 39.0 & 46.0 \\
Hybrid Risk Eval. (Alg.~1)       & 0.451 & 34.0 & 89.5 \\
Percentile + Hysteresis (Alg.~2) & 0.429 & 34.0 & \textbf{5.5} \\
\bottomrule
\end{tabular}
\end{table}

\subsection{Sensitivity to Hysteresis Parameters}
To study the responsiveness--stability trade-off, we vary the dwell window $N$
in Algorithm~\ref{alg:hysteresis}. Table~\ref{tab:hysteresis_sensitivity} shows
that increasing the dwell window $N$ consistently reduces switching frequency.
The effect on delay is moderate and not strictly monotonic. Settings around
$N=5$--$6$ provide a favorable trade-off between stability and latency, while
larger values ($N=8,10$) eliminate switching entirely for this trace.

\begin{table}[htpb]
\centering
\caption{Impact of the dwell window $N$ on switching frequency and delay}
\label{tab:hysteresis_sensitivity}
\renewcommand{\arraystretch}{1.1}
\setlength{\tabcolsep}{4pt}
\begin{tabular}{c c c c}
\toprule
\textbf{$N$} & \textbf{Switch Freq. (\%)} & \textbf{Mean Delay (s)} & \textbf{P95 Delay (s)} \\
\midrule
2  & 11.0 & 0.422 & 0.476 \\
3  &  9.5 & 0.426 & 0.476 \\
4  &  8.0 & 0.431 & 0.480 \\
5  &  5.5 & 0.429 & 0.476 \\
6  &  2.0 & 0.426 & 0.480 \\
8  &  0.0 & 0.414 & 0.472 \\
10 &  0.0 & 0.414 & 0.472 \\
\bottomrule
\end{tabular}
\end{table}

\subsection{Parameter Sensitivity and Practical Guidelines}
The percentile factor $k$ controls conservatism: larger values place more weight
on uncertainty and therefore improve protection against tail events, typically
with only a small effect on average latency. In practice, $k=1.645$,
corresponding to the one-sided 95th percentile under the Gaussian surrogate,
provides a robust default. Switching behavior is governed by the improvement
threshold $\Delta$ and the dwell window $N$. Smaller values favor responsiveness,
while larger values suppress oscillations. Our results indicate that moderate
settings around $\Delta=0.05$ and $N=5$--$6$ provide a favorable balance between
stability and latency. Across our traces, the Normal estimate and Cantelli bound
produced consistent server rankings, despite different absolute risk values.

\subsection{Scalability Considerations}
Although the real testbed uses three servers and a single client to isolate the
effect of the proposed decision logic, the framework is not inherently limited
to this scale. The per-decision complexity grows linearly with the number of
candidate servers and requires only simple arithmetic operations on predictive
summaries $(\mu_i,\sigma_i)$. No online training, global coordination, or
long-horizon optimization is required. Because decisions are made independently
per client and per frame, the framework is naturally compatible with
decentralized execution.

\subsection{Replay-Based Scalability Illustration}
\begin{figure}[t]
  \centering
 \includegraphics[width=1\linewidth]{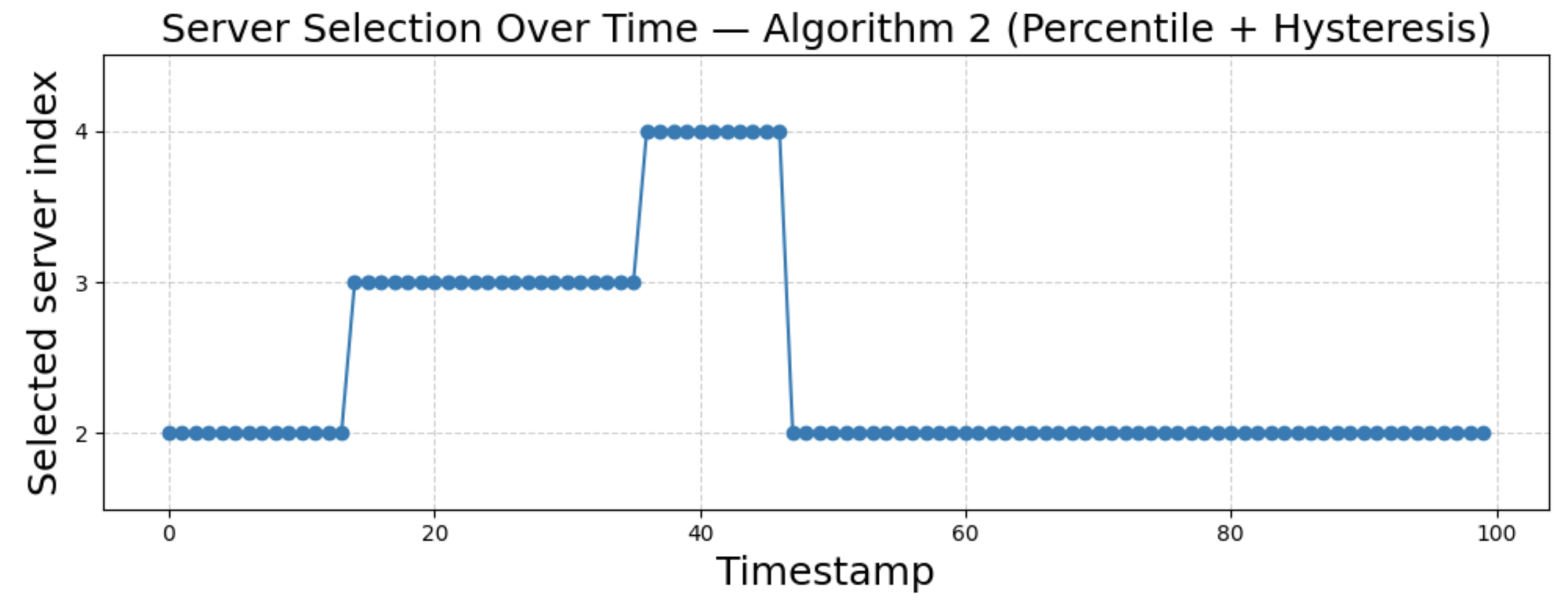}
  \caption{Selected server index over time under Algorithm~\ref{alg:hysteresis}
  in the containerlab-based replay with ten servers. The y-axis is restricted to
  the range of servers actually selected.}
  \label{fig:sim_algo2_10servers}
\end{figure}

To complement the real testbed and provide a broader-scale illustration, we replayed a containerlab-based scenario with ten candidate servers under the same decision logic.\footnote{\url{https://github.com/ldmohan/ContainerLab-IWCMC-2026}}
In this setting, latency summaries were more clearly separated and exhibited
limited short-term variability. Fig.~\ref{fig:sim_algo2_10servers} shows that,
although all servers are evaluated at each decision step, the policy converges
to a small subset of consistently low-risk candidates and maintains stable
operation with minimal switching.

This replay-based result does not replace large-scale real deployment, but it
supports the key computational property of the framework: per-decision overhead
grows linearly with the number of candidate servers and requires only simple
operations on $(\mu_i,\sigma_i)$.

\subsection{Remarks on Distributional Assumptions}
The Gaussian surrogate is used as a lightweight approximation for
percentile-based reasoning rather than as an exact model of network latency. In
practice, network delays may exhibit burstiness and heavier tails, especially
under contention. For this reason, the proposed framework combines the
Normal-based estimate with a one-sided Cantelli bound, which acts as a
conservative safeguard under distributional mismatch. Across our traces, both
measures produced consistent server rankings, suggesting that the decision logic
remains robust even when absolute risk values differ.

\subsection{Discussion}
Overall, the results show that incorporating uncertainty into server selection
reduces SLO violations by avoiding servers with unstable latency behavior, while
the hysteresis layer suppresses unnecessary handovers caused by short-lived
fluctuations. The framework therefore improves reliability and stability without
requiring fine-grained infrastructure visibility or learning-based control.

The experimental testbed in this paper is intentionally limited to three servers
and a single client to isolate the effect of the proposed decision logic. While
this scope is sufficient to validate the main design principle, broader
multi-client and more diverse workload scenarios remain important directions for
future evaluation.
\section{Conclusion and Future Work}
\label{sec:conclusions}

This paper presented a lightweight, network-centric decision framework for
dynamic edge server selection under latency SLOs. The proposed approach
combines probabilistic SLO-risk evaluation with percentile-based hysteresis to
transform uncertain network-delay summaries into reliable and stable server
selection decisions under time-varying conditions.

Experimental results on the real edge testbed showed that incorporating
uncertainty into server selection reduced the deadline-miss rate from 39\% to
34\% compared with a mean-only baseline. Moreover, the hysteresis mechanism
reduced switching frequency from 89.5\% to 5.5\% while maintaining the same DMR
and lowering the average delay to 0.429~s. These results demonstrate that
explicit risk evaluation combined with stability control can improve both
reliability and robustness without requiring internal server-state visibility or
learning-based control.

Future work will extend the framework to multi-client settings, mobility-aware
prediction, and broader real-world workload scenarios. An additional direction
is to investigate joint decision policies that incorporate latency, energy, and
cost objectives in federated edge deployments.

\section*{Acknowledgment}
This work was conducted as part of the project EMULATE, funded by the European Union and the German Federal Ministry of Economics and Climate Action under research grant 13IPC012.

\bibliographystyle{IEEEtran}
\bibliography{IWCMC_2026}
\end{document}